\documentclass[10pt,letterpaper]{article}
\usepackage{opex3}
\usepackage{epstopdf} 
\usepackage{cite}

\begin{document}

\title{Tracking of colloids close to contact}

\author{Chi Zhang, Georges Br\"{u}gger and Frank Scheffold$^*$}

\address{Department of Physics, University of Fribourg, Chemin du Mus\'{e}e 3, CH-1700, Fribourg, Switzerland}

\email{$^*$frank.scheffold@unifr.ch} 



\begin{abstract}
The precise tracking of micron sized colloidal particles - held in the vicinity of each other using optical tweezers - is an elegant way to gain information about the particle-particle pair interaction potential. The accuracy of the method, however, relies strongly on the tracking precision. Particularly the elimination of systematic errors in the position detection due to overlapping particle diffraction patterns remains a great challenge. Here we propose a template based particle finding algorithm that circumvents these problems by tracking only a fraction of the particle image that is insignificantly affected by nearby colloids. Under realistic experimental conditions we show that our algorithm significantly reduces systematic errors compared to standard tracking methods. Moreover our approach should in principle be applicable to almost arbitrary shaped particles as the template can be adapted to any geometry.
\end{abstract}



\section{Introduction}
Thanks to the pioneering work of Crocker et al.~\cite{Crocker1996} the combination of video microscopy and optical tweezers became a versatile tool to probe interaction potentials between colloidal particles ~\cite{Crocker1999} or to derive the rheological properties of the host material via two-point microrheology \cite{Crocker2P2000}. Since then the idea to analyse the relative trajectories of two colloids close enough to feel their interactions has been successfully applied to systems ranging from classical colloids \cite{Crocker1999, Crocker2P2000,Verma2000, Brunner2004, Baumgartl2005, Baumgartl2006, Biancaniello2006, Polin2008,Huang2009} to optically induced forces \cite{Brzobohaty2010, Baumgartl2009, Bruegger2015a} and bio-applications~\cite{Selhuberl2009, Gutsche2011,Roichman2015SM}.

The key point of the approach is to record statistical information about relative particle positions. When required optical tweezers can be used to keep the particles at a fixed average distance. To this end image sequences are acquired and analyzed with precise particle finding algorithms. Over the years, the continuous improvement of these algorithms concurrent with improved instrumentation has lead to a typical localization accuracy in the nanometer range for single particles and two-dimensional systems \cite{Biancaniello2006, Carter2005,fung2011measuring,Parthasarathy2012, Rogers2014}. However, one of the major limitations, the precise determination of the particles coordinates close to contact - when their diffraction patterns start to overlap - remains difficult and is still a field of ongoing research~\cite{Verma2000, Baumgartl2005, Baumgartl2006, Gutsche2011, Gyger2008}.

Avoiding systematic errors in the particle position determination at close distances is crucial for tracking based pair potential measurements. Baumgartl et al.~\cite{Baumgartl2005, Baumgartl2006} used numerically and experimentally obtained correction curves to account for the ill-determined particle positions. They could show that overlapping particle images can lead to an apparent attractive component in the particle pair potentials. In a different approach to overcome this problem van Blaaderen et al. used core-shell particles where only the core is labelled with a fluorescent dye~\cite{Blaaderen1995}. For a shell that is thick enough to prevent overlapping fluorescence patterns imaging artefacts can be avoided~\cite{Baumgartl2006, Blaaderen1995, Ramirez-Saito2006}. Crocker and co-workers~\cite{Verma2000, Biancaniello2006} proposed in an iterative procedure to subtract the image of one of the two particles. Eventually the position of the left over particle was determined using a conventional tracking algorithm.

As suggested by Gutsche et al.~\cite{Gutsche2011} we present in this paper a template based algorithm that only tracks the fraction of the particle whose optical pattern is barely altered upon the presence of a nearby second particle. The particles' images are considered as the convolution of the optical response of the imaging system with delta functions reflecting the location of the particles. Splitting the optical response into a residual part - that we ignore in the tracking process - and a part defined through the template allows for avoiding systematic tracking errors due to overlapping particle images. In a control experiment we show that our method suppresses tracking errors down to the noise level and thus increases the position accuracy substantially as compared to conventional tracking algorithms. 

\section{Materials}

\begin{figure}[ht]
\centering\includegraphics[width=9cm]{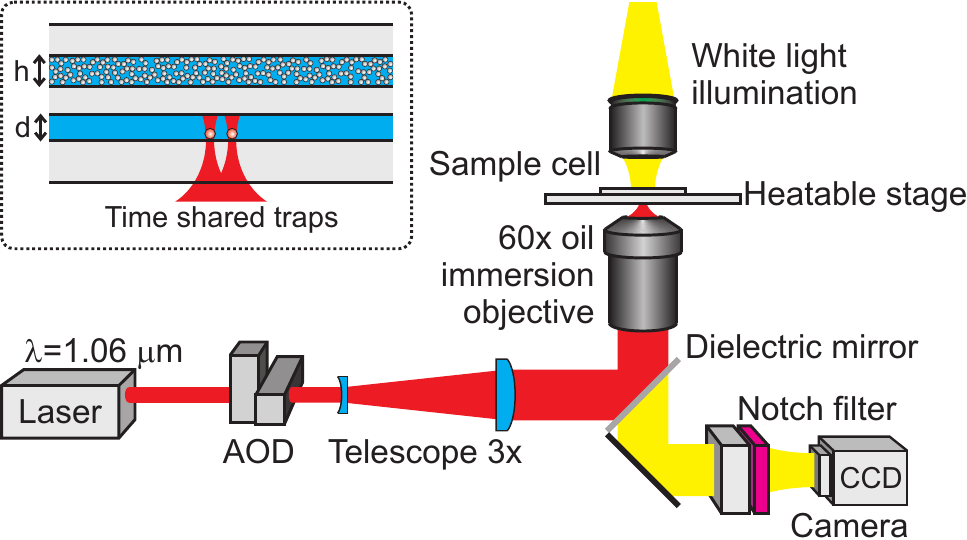}
\caption{The experimental setup is built around a commercial Nikon Eclipse TS100 microscope. A NIR laser operating at $\lambda=1064$~nm (shown in red) is used for the optical tweezing of the particles. Two acousto-optical deflectors (AOD) are employed to steer the laser beam at will. The white light illumination from above (depicted in yellow) is responsibele for the image acquisition. The inset shows a sketch of the
sample cell: The colloids are trapped at the bottom of a sandwich construction using the technique of time shared optical trapping~\cite{Sasaki1991, Fallman1997, Mio2000}. }
\label{fig:setup_AOD}
\end{figure}

We use an optical tweezers setup to investigate the effect of overlapping particle diffraction patterns on their measured center-to-center separation distance $r$, Fig. \ref{fig:setup_AOD}. For the trapping a near infrared laser beam (IPG Photonics, YLR-10-1064-LP) operating at a wavelength of 1064~nm and a power of $\sim$50~mW (measured at the exit of the laser) is expanded with a telescope (Thorlabs, 3x Galilean optical beam expander, BE03M-B) to optimize the trapping strength~\cite{Ashkin1992}. Subsequently the laser beam is fed into an oil immersion objective (Nikon 60x PlanApoVC, N.A.$=1.4$) and focussed into a sample cell to form an optical trap. The acousto-optic deflectors (AA Opto-Electronics) in the trapping beam path allow for steering the optical tweezers in $x$ and $y$ directions.
For the measurements melamine spheres with a diameter of 4~$\mu$m (microparticles GmbH, Germany) dispersed in a 3.3~mM KCl aqueous solution are confined between a capillary with a wall thickness of 20~$\mu$m (top) and a conventional cover slide (bottom) whose separation distance is controlled by adding a small amount of $d=15~\mu$m silica spheres. The capillary has a total thickness of $h=20~\mu$m (CM Scientific) and is filled with dense amorphous colloid composed of PMMA beads (Polymethylmethacrylat, diameter $\sim$400~nm) to form an opal diffuser. The whole sample cell is sealed with UV-curable glue.

Images of the colloidal particles are acquired using an adapted Nikon Eclipse TS100 bright field microscope equipped with a long working distance objective (20x/0.42 EO Plan Apo ELWD) for the white light illumination, an oil immersion objective (Nikon 60x PlanApoVC, N.A.$=1.4$) for both, the trapping and the observation of the particles motion using a CCD camera, an etch filter to filter the stray light coming from the trapping laser and a dielectric mirror to couple the trapping laser into the optical path of the microscope. Note, the bright field illumination takes place across the first diffusing layer allowing for an evenly white background illumination.

With the CCD camera (prosilica GC650) we record images of 200~$\times$~200 pixels with a frame rate of 90~Hz and an exposure time of 1.5~ms. With a micro-scale the effective pixel size is measured to be $d_{pix}\approx0.1~\mu$m. The recorded images are analysed using a custom made tracking algorithm derived in this work \cite{codefiletracking} to finally obtain the particle positions on each picture.

\section{Template based particle tracking method}
Consider a sample with $n$ identical particles. The image recorded by the imaging system can be written in terms of convolutions~\cite{B204414E,phdthesisBruijic2004}:
\begin{equation}\label{SSF}
  I({\bf r}) = \sum_{i=1}^n \delta({\bf r}-{\bf r}_{i}) \otimes S({\bf r})
\end{equation}
where ${\bf r}_{i}$ describes the position of the $i$-th particle and $S({\bf r})$ the response of the imaging system to the particle. Eq.~\ref{SSF} is accurate for dilute systems where individual particles are relatively far apart. Applying the convolution theorem to Eq.~\ref{SSF} allows for expressing particle locations in the following way~\cite{phdthesisBruijic2004}:
\begin{equation}\label{SSFb}
   \sum_{i=1}^n \delta({\bf r}-{\bf r}_{i}) = \mathcal{F}^{-1} \left[ \frac{\tilde{I}({\bf k})}{\tilde{S}({\bf k})} \right]
\end{equation} 
where 'tilde' denotes the Fourier transform and $\mathcal{F}^{-1}$ stands for the inverse Fourier transforms. The Fourier transform is very sensitive to noise and thus a filter is commonly applied to regularize the transformation. We closely follow the approach described in reference \cite{phdthesisBruijic2004} and use a Wiener filter $W =  \frac{|\tilde{S}|^2}{|\tilde{S}|^2+K}$ with $K = 10^7$ as a constant parameter \cite{russ2011image}. Here $|\tilde{S}|$ is the norm of the complex matrix $\tilde{S}$. Explicitly taking the filter into account in Eq. \ref{SSFb} the term on the right $\mathcal{F}^{-1}[\tilde{I}({\bf k})/\tilde{S}({\bf k})]$  can be rewritten as  $\mathcal{F}^{-1}[\tilde{I}({\bf k})/\tilde{S}({\bf k}) \times W]$. The same filter is used for all subsequent image analysis.
\begin{figure}[ht]
\centering\includegraphics[width=12cm]{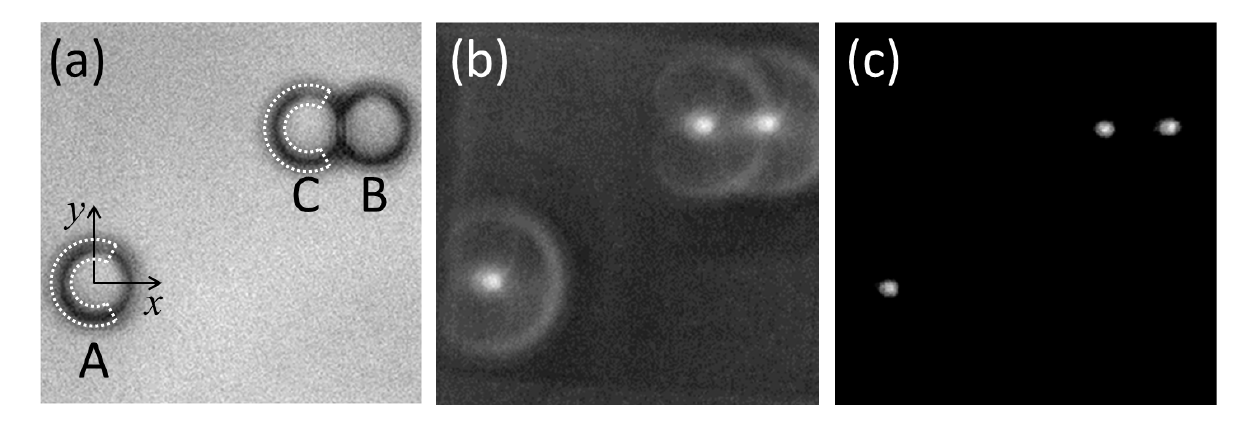}
\caption{Illustration of our template based tracking algorithm. (a) Image to be analyzed. The dashed lines illustrate the C-shaped particle regions that we use for tracking. (b) Backtransformed Fourier-convolved image according to Eq.~\ref{Ctempb} with a C-shape orientation $\alpha_C$ optimal for the localisation of particle C (and equally suited for A). The bright spots correspond to the particle locations and reflect the first term in Eq.~\ref{Ctempb}. The fuzzy shaped coronas surrounding the spots illustrate the second sum in Eq.~\ref{Ctempb}. (c) By choosing an appropriate threshold one can discriminate between particle locations and fuzzy coronas which allows for an precise and unbiased localisation of particle C. The same procedure is repeated to localize particle B using $\alpha_B= \alpha_C-\pi$.}
\label{fig:tracking_illustration}
\end{figure}
The standard centroiding based tracking approach breaks down for two particles close to contact. Overlapping diffraction patterns lead to ill-detected locations \cite{Baumgartl2005}. To circumvent these problems we use C-shaped templates and only track particle areas that are barely influenced by overlapping particle images. We proceed as follow:  the template matrix and the original image must have the same resolution. Taking this into account a ring is defined $u^2  < (x-x_0)^2+(y-y_0)^2 < (u+d)^2$ where $u$ is the inner radius  and $d$ is the ring thickness. $[x_0, y_0]$ denotes the center of the ring. The intensity inside the ring is set constant with a value given by the mean. The background outside the ring is filled with white noise  with a much lower mean value compared to the ring in order to avoid overflow in the fast Fourier transform. The "C"-shape is then created by discarding the elements within certain range of angles  $[-1/3 \pi, 1/3 \pi]$ from the tracking analysis and choosing an appropriate ring thickness $d\simeq 0.3 R$. We note that the choice of $d$ and of the opening angle of the C-shape is a tradeoff between optimizing the measured signal and effectively excluding the overlap area from the analysis.
\newline Next we divide the response of the imaging system $S({\bf r})=S_C({\bf r},\alpha)+S_L({\bf r},\alpha)$ into a C-shaped part $S_C({\bf r},\alpha)$ and a leftover part $S_L({\bf r},\alpha)$ where $\alpha$ defines the orientation of the C-shaped template. Using this notation Eq.~\ref{SSF} can now be written as
\begin{equation}\label{Ctemp}
  I({\bf r}) = \sum_{i=1}^{n}   \delta({\bf r}-{\bf r}_{i}) \otimes S_{C}({\bf r}, \alpha) + 
  			   \sum_{i=1}^{n}   \delta({\bf r}-{\bf r}_{i}) \otimes S_{L}({\bf r}, \alpha).
\end{equation}
To avoid systematic tracking errors caused by overlapping particle diffraction pattern, the proper orientation $\alpha$ of the C-shapes has to be found. In a first iteration one chooses an orientation $\alpha$ applicable to a certain particle or a subset of particles $n$. For the case of tracking a pair of particles this task is very simple: $\alpha$ can be derived by orienting the C-shaped template symmetrically to the line connection the centres of the particles, which is not affected by the tracking errors discussed here. The open side of the C-shapes have to face each other and the position of both particles has to be anaylzed separately using the appropriate value of $\alpha_C$ and $\alpha_C-\pi$. Subsequently the particle locations are found in analogy to Eq.~\ref{SSFb}:
\begin{equation}\label{Ctempb}
   \mathcal{F}^{-1} \left[ \frac{\tilde{I}({\bf k})}{\tilde{S}_{C}({\bf k, \alpha})} \right] 
    = \sum_{i=1}^{n} \delta({\bf r}-{\bf r}_{i}) \\ \nonumber
    + \sum_{i=1}^{n} \mathcal{F}^{-1} \left[\frac{\tilde{S}_{L}({\bf k})}{\tilde{S}_{C}({\bf k}, \alpha)} \right] \delta({\bf r}-{\bf r}_{i}).
\end{equation}
The procedure is illustrated in Fig.~\ref{fig:tracking_illustration}(a) showing a sample image of $n=3$ identical particles acquired in the experiment described in the Results section. In a first iteration we chose a C-template $S_C({\bf r},\alpha_C)$ with an orientation $\alpha$ appropriate to localize the particle on the left side and thus obtain the image shown Fig.~\ref{fig:tracking_illustration}(b); a graphical representation of Eq.~\ref{Ctempb}. The bright spots correspond to the particle locations, i.e., they visualize the first sum in Eq.~\ref{Ctempb}. Additionally, we observe white coronas surrounding the bright spots. These patterns - referring to the second sum of Eq.~\ref{Ctempb} - compromise the tracking accuracy. We eliminate them by setting an appropriate lower threshold [see Fig.~\ref{fig:tracking_illustration}(c)]. Finally we determine the particle coordinates by fitting the pixel intensities of the white peaks depicted in Fig.~\ref{fig:tracking_illustration}(c) to a Gaussian. We now turn our attention to particle B: It can be seen that the intensity peak for particle B is slightly skewed with respect to the peaks for the other two particles. This is because in Fig.~\ref{fig:tracking_illustration}(b) the template $S_C({\bf r},\alpha_C)$ is adapted for particle C (and trivially to A as well). To find the coordinates of particle B we repeat in a second iteration the procedure from above with a new template $S_C({\bf r},\alpha_B)$ with an orientation $\alpha_B=\alpha_C-\pi$ adjusted to particle B (\protect\rotatebox[origin=c]{180}{C}-template). 
Finally we must note that due to the asymmetry of the C-template the measured coordinates  are slightly shifted compared to their actual positions. This small offset can be accounted for by applying the tracking algorithm to an isolated particle in the same field of view as a reference (our particle A). In the situation depicted in Fig.~\ref{fig:tracking_illustration}(a), the position of particle A is detected via the C-templates first (both for $\alpha_C$ and $\alpha_C-\pi$). Comparing the so obtained coordinates to the accurate ones determined with the standard tracking approach applied to particle A~\cite{Crocker1996, phdthesisBruijic2004,gao2009accurate} allows for quantifying the offset precisely. 
\newline It is noteworthy that the presented tracking method is not restricted to C-templates only. In principle the templates could be adjusted to track almost arbitrary shaped particles. Also the orientation of the templates can be adapted according to the symmetry of the particle assembly and could be generalized to three or more particles. An implementation of other template shapes and the experimental application to non-spherical particles is however beyond the scope of this work. 

\section{Results}

\begin{figure}[ht]
\centering\includegraphics[width=9cm]{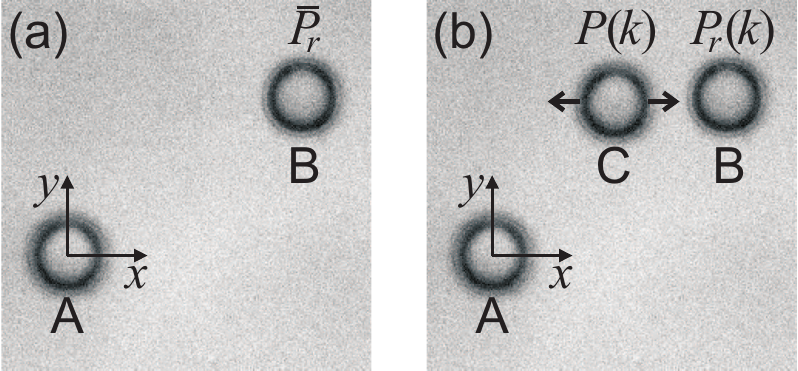}
\caption{Illustration of the experimental procedure to quantify position errors due to a certain tracking algorithm. (a) Two isolated particles are used to determine the true position $\bar{P}_r$ of particle B relative to A. (b) In a subsequent experiment a third particle C is trapped in the vicinity of particle B and its position periodically scanned. If particles C and B are close enough for their diffraction patterns to overlap conventional tracking algorithms start to fail and the instantaneous position $P_r(k)$ deviates form the true position $\bar{P}_r$.}
\label{fig:error_explanation}
\end{figure}

To study the effect of two overlapping particle diffraction patterns we first locate two particles that are irreversibly stuck to the glass substrate within the field of view of the camera as depicted in Fig.~\ref{fig:error_explanation}(a). The screening of the mutual electrostatic repulsion between the colloids and the glass surface by the addition of salt simplifies this task and leads to a situation where after one day a significant amount of the melamine particles are irreversibly adsorbed to the glass substrate. In a first experiment we then determine form a movie of 10'000 images the mean position $\bar{P}_r=(\bar{x}_r,\bar{y}_r)$ of particle B relative to particle A using our and a conventional tracking algorithm~\cite{Crocker1996}. We use particle A to define the origin of the coordinate system to avoid drift effects that possibly could spoil the accuracy of the tracking algorithms.
As illustrated in Fig.~\ref{fig:error_explanation}(b), in a second experiment we record another sequence of 100'000 images where a previously freely moving colloid (particle C) is trapped in the vicinity of particle B and its position periodically scanned in $x$-direction using the acousto-optic deflectors. Subsequently we determine on each image $k$ the instantaneous location of particle B, $P_r(k)=[x_{r}(k),y_{r}(k)]$, and particle C, $P(k)=[x(k),y(k)]$, in the coordinate system set by particle A. For a correct tracking we expect $P_r(k)$ to be independent on the position of the trapped particle $P(k)$, i.e., $\langle [x_r(k),y_r(k)]\rangle=(\bar{x}_r,\bar{y}_r)$. As pointed out by Refs.~\cite{Baumgartl2005, Baumgartl2006} this implies that one possible quantitative measure of how overlapping particle diffraction pattern affect a correct trapping is obtained by plotting $\Delta r(k)=r(k)-\sqrt{[x(k)-\bar{x}_r]^2+[y(k)-\bar{y}_r]^2}$ as a function of the instantaneous separation distance between particle C and B, $r(k)=\sqrt{[x_r(k)-x(k)]^2+[y_r(k)-y(k)]^2}$.

\begin{figure}[ht]
\centering\includegraphics[width=12cm]{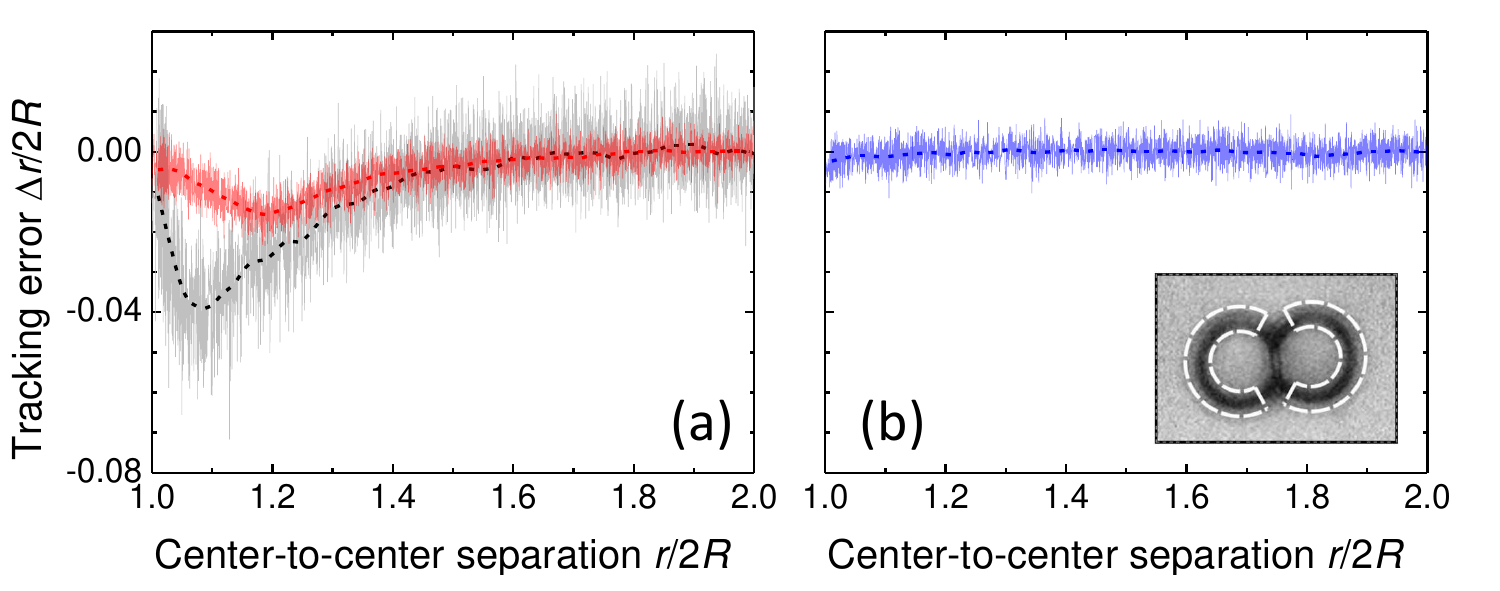}
\caption{Tracking errors $\Delta r/(2R)$ vs. instantaneous separation distance $r$ are shown as solid lines ~\cite{Baumgartl2005, Baumgartl2006}. Dashed lines show the average values. (a) Error estimation for the particle tracking data obtained with the conventional Crocker and Grier~\cite{Crocker1996} tracking algorithm (grey line) and with the more accurate SSF-refinement technique method \cite{jenkins2008confocal} (red line). (b) Error according to our template based tracking method.  The inset illustrates the C-shaped particle regions that we use for tracking. The near-contact area where interference effects are the most pronounced is excluded form the tracking analysis.}
\label{fig:tracking_error}
\end{figure}

In Fig.~\ref{fig:tracking_error} (a) and ~\ref{fig:tracking_error}(b) we plot the tracking error $\Delta r$ as a function of the instantaneous separation distance $r$ between particle C and B in units of the particle diameter $2R$. Fig.~\ref{fig:tracking_error}(a) shows the result for the standard centroiding particle tracking algorithm according to Crocker and Grier~\cite{Crocker1996,polin2007colloidal}. At short separations the presence of a second particle has a clear effect on the correct position determination. Indicated by the negative $\Delta r$ values the instantaneous separation $r$ systematically underestimates the true separation distance with an error peak around 1.1 particle diameters. For comparison we also show the results of the refined spheres spread function (SSF) approach which provides a better accuracy but still suffers from a systematic positional shift toward the neighbouring particle as already noted in reference  ~\cite{jenkins2008confocal}.

In Fig.~\ref{fig:tracking_error}(b) we analyse the same set of images using our template based tracking algorithm. The main idea is to track a C-shaped fraction of the ring shadows that is only little affected by interference phenomena due to a nearby second particle. We are well aware of the fact that the presence of a close-by second particle changes the appearance of the entire particle. However, it turns out that tracking the outer part of the black rings significantly decreases the tracking errors $\Delta r$ over the whole range of separations $r$. 


\section{Conclusion}
Our template based algorithm allows for a precise position determination for particles close to contact. The tracking error due to overlapping diffraction patterns is reduced down to the noise level of the measurement. Therefore we expect the proposed method to simplify and improve interaction potential measurements based on particle tracking algorithms. It omits cumbersome sample synthesis steps and/or time consuming image post processing. Moreover, our results suggest that in future implementations such a template based tracking code could be easily adapted to track almost arbitrarily shaped particles or to handle three or more overlapping particle diffraction patterns. 
\section*{Acknowledgments
} This work was supported by the Swiss National Science Foundation through projects No.132736 and No.149867 and through the National Centre of Competence in Research (NCCR) \emph{Bio-Inspired Materials}.
\end{document}